
\input harvmac.tex

\def\mm{{_{=}}}
\def\pp{{_{+\!\!\!+}}}
\def\p{{_{+}}}
\def\m{{_{-}}}

\def\bu{b_1}
\def\cu{c_1}
\def\bd{\bar b_1}
\def\cd{\bar c_1}
\def\eu{\epsilon}
\def\ed{{\bar \epsilon}}

\def\nablapp{\nabla_{\!\!\pp}}
\def\nablamm{\nabla_{\!\!\mm}}
\def\chippp{\chi_{\p\!\pp}}
\def\chimmm{\chi_{\m\!\mm}}
\def\chimpp{\chi_{\m\!\pp}}
\def\chipmm{\chi_{\p\!\mm}}
\def\mnbox#1#2{\vcenter{\hrule \hbox{\vrule height#2in
                \kern#1in \vrule} \hrule}}  
\def\sq{\,\raise.5pt\hbox{$\mnbox{.09}{.09}$}\,}
\def\sqb{\,\raise.5pt\hbox{$\overline{\mnbox{.09}{.09}}$}\,}
\def\sqr#1#2{{\vcenter{\vbox{\hrule height.#2pt
     \hbox{\vrule width.#2pt height#1pt \kern#1pt
           \vrule width.#2pt}
       \hrule height.#2pt}}}}

\Title{\vbox{\baselineskip12pt\hbox{NBI-HE-93-69}
\hbox{hep-th/9311157} \hbox{November 1993}
}}
{{\vbox{\centerline{A Locally Supersymmetric Action}
\bigskip \centerline{ for the Bosonic String }}}}

\centerline{
Fiorenzo Bastianelli\footnote{$^\dagger$}{e-mail:
fiorenzo@nbivax.nbi.dk}}
\bigskip\centerline
{\it  The Niels Bohr Institute}
\centerline {\it  University of Copenhagen}
\centerline {\it Blegdamsvej 17}
\centerline {\it DK-2100 Copenhagen {\O}, Denmark}
\vskip 1in

\noindent
Recently Berkovits and Vafa have shown that the bosonic string
can be viewed as the fermionic string propagating in a particular
background. Such a background is described by a somewhat unusual
$N=1$ superconformal system. By coupling it to $N=1$ supergravity
I construct a local supersymmetric action for the bosonic string.
\vskip .8cm

\Date{ }
\eject

In a recent paper Berkovits and Vafa \ref\BV{N. Berkovits
and C. Vafa, \lq\lq On The Uniqueness of String Theory",
preprint HUTP-93/A031, KCL-TH 93-13, hep-th/9310170 (October 1993).}
have shown how to embed the bosonic string into the $N=1$
superstring. After identifying a suitable  $N=1$ superconformal
system which could be used as a background for the $N=1$
string, they showed how the bosonic string amplitudes are reproduced
by the fermionic string propagating on it.
They also showed how to view the $N=1$ string
as the $N=2$ string moving in a particular background, and were
led to conjecture the existence of an \lq\lq universal string theory"
which contains all string theories as particular choices
of the vacuum (see also ref. \ref\jose{J.M. Figueroa-O'Farrill
\lq\lq On the Universal String Theory",
preprint QMW-PH-93-29, hep-th/9310200 (October 1993).}
 for some interesting remarks on the
search for an universal string theory).
In this letter I analyze the structure of the $N=1$
superconformal system used by Berkovits and Vafa to embed the bosonic
string into the $N=1$ string. I look at the classical limit
of such a system and by coupling it to $N=1$ supergravity
I shall produce a locally supersymmetric action for the bosonic string.
I will conclude presenting few comments on its quantization
and on its relation to the usual bosonic string action.

I start by recalling the Berkovits-Vafa superconformal system.
It is realized on a matter system with stress tensor $T_m$
and central charge $c=26$ plus an anticommuting bc-system
$(b_1,c_1)$ of spin $({3\over 2},- {1\over 2})$.
The generator of the $N=1$ superconformal algebra are
\eqn\gen{\eqalign { &G=
\bu + \cu( T_m +\partial \cu \bu )
+ {5\over 2} \partial^2 \cu \cr
&T =  T_m - {3\over 2} \bu \partial \cu - {1\over 2} \partial
\bu \cu + {1\over 2}\partial^2 (\cu\partial\cu),
\cr}}
and their OPE generates a superconformal algebra with $c=15$.
For simplicity I will  consider the matter system as given by
the usual 26 free scalar fields $X^i$, and
omit showing the index $i$ in the following.
By dropping the improvement terms in \gen, i.e. considering
the generators
\eqn\cgen{\eqalign { &G_0 =
\bu + \cu( -{1\over 2} (\partial X )^2 +\partial \cu \bu ) \cr
&T_0 =  -{1\over 2}(\partial X)^2
  - {3\over 2} \bu \partial \cu - {1\over 2} \partial
\bu \cu,  \cr}}
and keeping only single contractions in their OPE, one obtains
a classical superconformal algebra without central charges.
Such a superconformal algebra, together with its right-moving
counterpart, describes the classical symmetries of the action
\eqn\freeact{ S= {1\over \pi} \int \! d^2x \
[{1\over 2} \partial X \bar \partial X + \bu \bar \partial \cu +
\bd \partial \cd ] . }
The supersymmetry
transformation rules can be easily computed from the generators
and read
\eqn\clatr{\eqalign{
 \delta X &= \eu \cu \partial X +
\ed \cd \bar \partial X , \cr
\delta \cu &=  \eu( 1+ \cu \partial \cu), \ \ \ \ \ \ \
\delta \bu =  \partial (\eu \cu \bu)
-\eu \bigl ( {1\over 2} (\partial X)^2
+ \bu \partial \cu \bigr ), \cr
\delta \cd &=  \ed( 1+ \cd \bar \partial \cd), \ \ \ \ \ \ \
\delta \bd =  \bar \partial (\ed \cd \bd)
-\ed \bigl ( {1\over 2} (\bar \partial X)^2
+ \bd \bar \partial \cd \bigr ), \cr  }}
where $\bar \partial \eu =0$ and $\partial \ed =0$.
This realization of supersymmetry is quite unusual.
It looks spontaneously broken ($\delta \cu = \eu + \dots$)
and it is non-linearly realized. For this last reason
it is difficult to rewrite the model with
superfields and couple it to supergravity
using standard superspace techniques.
However, one can follow another well-known path, that of using
the Noether method to gauge global symmetries. This
was in fact the strategy used in \ref\paolo{L. Brink,
P. Di Vecchia and P. Howe, Phys. Lett. B65 (1976) 471\semi
S. Deser and B. Zumino, Phys. Lett. B65 (1796) 369.}
to construct a locally supersymmetric action for the spinning
string. I will now follow the same path for the Berkovits-Vafa
supersymmetric model and  obtain
 a locally supersymmetric action for the bosonic string.
The coupling to gravity is straightforward. One introduces the
vielbein $e_\mu{}^a$ and Lorentz covariant derivatives
$\nabla_a = e_a{}^\mu \partial_\mu + \omega_a J$, where the
flat index $a$ takes the values
$(+\!\!\! +,=)$, $J$ is the Lorentz generator which measures
the Lorentz spin, and $\omega_a$ is the spin connection.
These derivatives satisfy the relation
$[\nablamm , \nablapp] =  R J$,
with $R$  the curvature scalar\footnote{$^1$}{The
Lorentz metric is: $\eta_{\pp \mm}= \eta_{\mm \pp} =
{1\over 2}$, $\eta_{ab}= 0$ otherwise. The notation used here
has been fully described in appendix A of ref.
\ref\me{F. Bastianelli and U. Lindstr\"om, \lq \lq Induced
Chiral Supergravities in 2D", preprint USITP-93-04, hep-th/9303109,
Nucl. Phys. B to appear.}.}.
Gauging the supersymmetry requires the introduction of the gravitino
fields $ \chippp =\chi_{\p\mu} e^\mu{}_\pp$ and
$\chimmm =\chi_{\m \mu} e^\mu{}_\mm$. The Noether method
shows then the need of a linear coupling to the supercurrents as well
as a term quadratic in the gravitino.
The invariant action is
\eqn\action{\eqalign{
S= {1\over \pi} \int \! d^2x\ e & \bigl [
{1\over 2} \nablapp X \nablamm X + \bu \nablamm \cu +
\bd \nablapp \cd  + \chimmm G_{\p\!\pp} + \chippp G_{\m\!\mm}
\cr &
+ \chippp \chimmm \cu\cd \nablapp X \nablamm X \bigr ] \cr }}
where
\eqn\defi{\eqalign{ G_{\p\!\pp}&=
\bu (1 + \cu \nablapp \cu) - {1\over 2} \cu
(\nablapp X)^2 \cr  G_{\m\!\mm}&=
\bd (1 +  \cd \nablamm \cd) - {1\over 2} \cd (\nablamm X)^2,
 \cr}}
and where $ (\bu , \cu , \bd ,\cd ) $ are Lorentz tensors with spin
$({3\over 2},- {1\over 2}, - {3 \over 2}, {1\over 2})$.
It is manifestly reparametrization invariant, and  the supersymmetry
transformation rules on the matter fields are as follows
\eqn\symm{ \eqalign{
\delta X &= (\eu -\ed \cd \chimmm) \cu \nablapp X +
            (\ed -\eu \cu \chippp) \cd \nablamm X   \cr
\delta \cu &=  \eu( 1+ \cu \nablapp \cu) +\eu \chi_{\m\pp} \cu
\cr
\delta \cd &=  \ed( 1+ \cd \nablamm \cd) +\ed \chi_{\p\mm} \cd  \cr
\delta \bu &=  \nablapp (\eu \cu \bu)
-\eu \bigl ( {1\over 2} (\nablapp X)^2
+ \bu \nablapp \cu \bigr )
-\eu \cd \chippp \nablapp X \nablamm X
 \cr & \ \ \ \
-3\eu \chimpp \bu
+ \chimmm \chippp \bigl(
\cu\cd \eu (\nablapp X)^2  - 2 \ed \bu \cu \bigr )
\cr
\delta \bd &=  \nablamm (\ed \cd \bd)
 -\ed \bigl ( {1\over 2} (\nablamm X)^2
+ \bd \nablamm \cd \bigr )
-\ed \cu \chimmm \nablamm X \nablapp X
\cr & \ \ \ \
 -3\ed \chipmm \bd
+ \chippp \chimmm \bigl (
\cd\cu \ed (\nablamm X)^2  - 2 \eu \bd \cd \bigr ).
\cr
}}
The transformation rules for the supergravity multiplet
are the standard ones, written here
without the use of the gamma matrices
\eqn\prova{ \eqalign{
\delta e_\mu{}^\pp &= 2 \eu \chi_{\m \mu} \cr
\delta e_\mu{}^\mm &= 2 \ed \chi_{\p \mu} \cr
\delta \chi_{\m\mu} &= \nabla_{\!\mu} \eu + ( \chi_{\m \mu}
\chimpp - \chi_{\p \mu} \chipmm) \eu  \cr
\delta \chi_{\p\mu} &= \nabla_{\!\mu} \ed + ( \chi_{\p \mu}
\chipmm - \chi_{\m \mu} \chimpp) \ed . \cr}}
Alternatively, one can deduce the transformation rules
for the gravitino with flat indices, which
appears  directly in the
action
\eqn\alt{\eqalign{
\delta \chimmm &= \nablamm \eu + \eu \chimpp \chimmm -2 \ed
\chipmm \chimmm ,
\ \ \ \ \ \
\delta \chimpp =\nablapp \eu +\eu \chipmm \chippp
-2\ed \chippp \chimmm,   \cr
\delta \chippp &= \nablapp \ed + \ed \chipmm \chippp -2 \eu
\chimpp \chippp,
\ \ \ \ \ \
\delta \chipmm =\nablamm \ed +\ed \chimpp \chimmm
-2\eu \chimmm \chippp .  \cr}}
Note that the action is super-Weyl invariant since it is manifestly
independent of the fields $\chimpp$ and $\chipmm$.

One may use  the action \action\ in a Polyakov approach
to compute string amplitudes.
When quantizing the matters fields $(X,\bu,\cu,\bd,\cd)$
in the supergravitational background, one discovers anomalies
that, however, can be canceled by counterterms.
This can be seen most easily by first choosing the conformal gauge
and then checking that the anomalies, which now appear in the BRST algebra,
are canceled by counterterms.
Let's consider for example the embedding of the bosonic string
into the $N=(0,1)$ heterotic string. The corresponding $N=(0,1)$
locally supersymmetric action  is
obtained by setting the fields  $\chi_{\p\mu}= \bd=\cd =0$ in
\action\ and \symm. I quantize it by using
 the Batalin-Vilkovisky method for lagrangian quantization
\ref\bat{ I.A. Batalin and G.A. Vilkovisky, Phys. Lett. B102 (1981)
27.} (for a review see ref. \ref\toine{ A. Van Proeyen, in \lq\lq
String and Symmetries 1991", eds. N. Berkovits et al.
(World Scientific, Singapore, 1992).}). I choose the conformal gauge
and  obtain the following gauge-fixed action\footnote{$^2$}{For
simplicity, I consider here only the relevant (supersymmetric) chiral
sector.}
\eqn\gfaction{ \eqalign{ S= {1\over \pi} & \int \! d^2x \
\bigl [
{1\over 2} \partial X \bar \partial X + \bu \bar \partial \cu
+ b \bar \partial c + \beta \bar \partial \gamma +
X^* ( c - \gamma \cu) \partial X \cr &
+ \cu^* ( - c\partial \cu
+{1\over 2} \partial c \cu +\gamma + \gamma \cu \partial \cu)
\cr &
+ \bu^* \bigl ( -c \partial \bu - {3\over 2} \partial c \bu +
\partial ( \gamma \cu \bu) -
\gamma ({1\over 2} (\partial X)^2 + \bu \partial \cu )
\bigr )\cr &
+ c^* ( -c \partial c + \gamma^2)
+ \gamma^* (c \partial \gamma - {1\over  2} \gamma \partial c )
- b^* ( T_0 + T_{gh})
- \beta^* ( G_0 + G_{gh} ) \bigr ]
\cr}}
where $(b, c)$ and $(\beta, \gamma)$ are the ghosts for the
reparametrization and local supersymmetry,
$T_0$ and $G_0$ are given in \cgen, and
\eqn\ghostcurr{\eqalign {
T_{gh} &= -2 b\partial c - \partial b c - {3\over 2} \beta \partial \gamma
-{1\over 2} \partial \beta \gamma \cr
G_{gh} &= -c\partial \beta - {3\over 2} \partial c \beta + 2 \gamma b .
\cr}}
The starred fields, called antifields, are sources for
the BRST variations.  The gauge-fixed action in
\gfaction\ satisfies the classical master equation of
Batalin-Vilkovisky\footnote{$^3$}{The definitions of the
antibracket $(\ ,\ )$ and of the operator $\Delta$ appearing in the
quantum master equation are standard. See refs. \bat,\toine.}:
 $(S,S) =0$.
However, at the quantum level it must satisfy the
quantum master equation: $ (S,S) = 2 \Delta S $.
One can think of $\Delta$ as to the operator which computes
anomalies in the BRST symmetry. The quantum master equation can be
satisfied if such anomalies are canceled  by the BRST variation
of appropriate counterterms (which are then included in the action
appearing in the quantum master equation).
In our example this is achieved by adding to \gfaction\ the
following counterterms\footnote{$^4$}{In computing loops and anomalies
one needs to specify a regularization scheme. Here I use analyticity
in configuration space,
as standard in two dimensional conformal field theory.}
\eqn\ctr{ S_{ct}= {1\over \pi} \int \! d^2 x\  \bigr [
\bu^* ( {1\over 2} \partial^3 c \cu +\partial^2 c \partial \cu
+{5\over 2} \partial^2 \gamma )
- b^* {1\over 2} \partial^2 ( \cu \partial \cu) -\beta^* {5\over 2}
\partial^2 \cu \bigl ]}
These counterterms describe nothing else than the improvement terms
present in \gen, needed to close the BRST quantum algebra.

Now I comment on the relation of the action \gfaction\
to the usual action for the bosonic
string, again employing the Batalin-Vilkovisky formalism.
In such a formalism canonical transformations are typically used
to gauge-fix and to redefine
variables (in fact, the process of gauge-fixing can be thought
of as a particular field redefinition). Canonical transformations
are specified by a fermionic generating function
$\Psi$ and are defined by $ \phi \rightarrow \phi' =
{\rm e}^{{\cal L}_\Psi} \phi$, where $\phi$
is a field or an antifield,
and $ {\cal L}_\Psi \phi \equiv (\Psi,\phi)$.
Performing a canonical transformation on \gfaction\ with
\eqn\canuno{\Psi_1 = \int\! d^2x\ [\beta^* b\cu + \bu^* b \gamma -
 c^* \gamma \cu ]}
redefines variables so that the coordinates $X$ become inert under
\lq\lq supersymmetry".
A second canonical transformation generated by
\eqn\candue{ \Psi_2 = \int \! d^2x\  {1\over2} \bigr [
 \beta^* \beta \cu \partial \cu
-\gamma^* \gamma \cu \partial \cu
 - \bu^* \bigl (\partial (\beta \gamma \cu)
+\beta \gamma \partial \cu \bigr ) \bigr ]}
simplifies the BRST transformation rules
 in the $(\bu,\cu ,\beta ,\gamma)$
sector and a final canonical transformation generated by
\eqn\cantre{ \Psi_3 = \int \! d^2x\ \bigl [ \gamma^* (-c\partial \cu
+{1\over 2} \partial c \cu) - \bu^* ( c\partial \beta +
{3\over 2} \partial c \beta )- b^* ({1\over 2} \cu \partial \beta
+ {3\over 2} \partial \cu \beta ) \bigr ] }
achieves the  decoupling of the BRST \lq\lq
reparametrizations" rules from such a sector. One is left with
the following action
\eqn\gfa{ \eqalign{ S= {1\over \pi} & \int \! d^2x \
\bigl [{1\over 2} \partial X \bar \partial X + b \bar \partial c +
\bu \bar \partial \cu  + \beta \bar \partial \gamma \cr & +
X^* c  \partial X
+ c^* (-c\partial c)
- b^* \bigl ( -{1\over 2} (\partial X)^2 -2b\partial c - \partial b
c \bigr ) +\cu^* \gamma - \beta^* \bu \bigr ] \cr}}
which  shows that the fields $(\cu, \gamma)$
and $( \beta, \bu)$ are just non-minimal fields \bat,\toine.
They can be dropped for free, since they carry no BRST cohomology.
This can be easily seen  by performing an additional canonical
transformation generated by
 $\Psi_4= \int \! d^2 x \beta ( \cu - \bar \partial
\cu) $ to  simplify their kinetic term.
Now they can be trivially eliminated, and
one is left with the standard gauge-fixed action for the bosonic
string. What I have just described
is a different way of viewing the equivalence (though at the
classical level) between the $N=1$ string propagating on  the Berkovits-Vafa
type of background and the usual bosonic string.
The quantum version of the canonical transformation here described
has been also found in a recent paper by Ishikawa and Kato
\ref\IK{ I. Ishikawa and M. Kato, \lq\lq Note on
the $N=0$ string as the $N=1$ string",
preprint UT-Komaba/93-23, hep-th/9311139 (November 1993).}.

\bigskip

{\bf Acknowledgments}

It is a pleasure to thank Nobuyoshi Ohta
for discussions and valuable comments.

\listrefs
\end